\documentstyle[prb,aps]{revtex}
\begin {document}
\input{psfig}
\bibliographystyle {plain}
\title
{\bf Skyrmions in the quantum Hall effect at finite Zeeman coupling.}
\author{A. G. Green, I. I. Kogan  and A. M. Tsvelik}
\address{Department of  Physics, University of Oxford, 1 Keble Road,
Oxford OX1,  UK} 

\maketitle

\begin{abstract}
Using a self consistent approximation for the spin distribution of 
Skyrmions 
in the quantum Hall effect, we obtain an effective action for the 
Skyrmion 
coordinates. The energy functional is minimised for a periodic 
distribution 
of Skyrmions with an underlying hexagonal symetry. We calculate the 
phonon 
spectrum of this lattice and find that near the classical minimum, 
breathing 
modes of the Skyrmions are strongly suppressed. The resulting equal sized 
Skyrmions interact via a residual Coulomb potential. Neglecting coupling 
between phonons and spin waves due to dipole and higher order Coulomb 
interactions, the Skyrmion crystal has a phonon spectrum identical to 
that of 
an electronic Wigner crystal. The transition to a liquid of equal sized 
Skyrmions is discussed using the theory of dislocation melting and a 
comparison is made between the predictions of this theory and the results 
of 
a recent experiment.  

\end{abstract}
\pacs{PACS numbers 73.40. H}

\section{Introduction.}

The theoretical prediction of Skyrmion spin textures in the fractional
quantum Hall effect, made by Sondhi et al. \cite{sondhi}, has
prompted much theoretical work, supported by mounting
experimental evidence, (see Refs. [\onlinecite{barrett,aifer,schmeller}]).

Here, we consider the low temperature properties of a multi-Skyrmion system.
 Several parallel approaches have been developed to solve this problem. 
Historically, the first
method, used in Ref.[\onlinecite{sondhi}], employed a field theoretic
description using a non-linear sigma model with a topological term. A 
solution for the partition function of
topological excitiations or Skyrmions in this model, in the absence of
Zeeman and Coulomb interactions, was found by Fateev {\it et
al}. \cite{fateev}. At zero Zeeman energies and for
sufficiently high temperature, this solution applies directly to the
quantum Hall effect \cite{green} and describes a meron liquid. 
When the
Zeeman energy is finite, the problem is complicated by the interplay
between the Coulomb and Zeeman energies. As originallly pointed out by
Sondhi et al.\cite{sondhi}, this fixes the scale for a single 
Skyrmion. In the
multi-Skyrmion case, the problem is complicated further by the
non-local nature of a Skyrmion spin distribution. The common belief, 
based on the analysis of a single Skyrmion, is that we have equal sized 
Skyrmions with some kind of residual
interaction. We shall demonstrate that this belief is justified in the 
physically interesting range of parameters.

An alternative approach is to use a Hartree-Fock
approximation\cite{fertig}. 
In particular
L. Brey {\it et al}. in Ref.[\onlinecite{brey}] suggest that the zero
temperature ground state of a multi-skyrmion system is a periodic
structure rather like a Wigner crystal and pick out a face centred 
square lattice
as the most stable of these. Our results, however, indicate that 
the ground state is closer to (or even perhaps, coincides with) 
a hexagonal structure.

Recently, a full microscopic quantum mechanical solution of the
problem with a delta function interaction was presented by MacDonald {\it et
al}. \cite{macdonald}. This approach requires computer 
diagonalization of the truncated Hamiltonian. It seems that it 
is rather difficult to apply this approach to systems with  
large numbers of Skyrmions. 

 In this paper we continue to develop the field theoretic description of the
problem. In Section II, we introduce the energy functional and
describe  the Skyrmion solution. In Section III, we argue that for
small Zeeman coupling the Skyrmions are only slightly distorted by
the Coulomb and Zeeman interactions and are well approximated by the
Belavin-Polyakov solutions \cite{belavin}. 
We use this solution to investigate the experimentally relevant  
regime  of Skyrmions of a small size.
 In Section V, we
show that near the classical energy minimum the Skyrmions all have equal size
and interact via a residual point-like Coulomb interaction. At low 
temperatures a Wigner crystal with rectangular symetry is formed. We give an
analytic expression for this structure and highlight its relationship
to the square lattice predicted by Brey {\it et al}. and the hexagonal 
lattice expected for an electron Wigner crystal. In section VI, the
phonon spectrum of the Skyrmion lattice is calculated followed by the 
spinwave spectrum in Section VII. In Section VIII, we discus the
dislocation mediated melting of the Skyrmion crystal to form a liquid
of equal sized Skyrmions. The results of this analysis are compared
waith recent experiment. Finally in Section
IX, we present a summary of our main results.

\section{Skyrmions in the quantum Hall effect.}

\subsection{The Energy Functional.}
In ref.[\onlinecite{sondhi}], the following energy functional is 
suggested for the
ferromagnetic order parameter in the fractional quantum Hall effect:
\begin{eqnarray}
E &=& E_{nl\sigma}+E_{Zeeman}+E_{Coulomb} \nonumber\\
&=& \int d^2x \left[ \frac{\rho_s}{2}|\partial_{\mu}{\bf n}|^2
+\frac{\mu q}{\nu}- g_L\bar \rho B n^z  \right]  
+\frac{1}{2} \int d^2x \int d^2x'q(r)V(r - r')q(r'),\label{energy}
\end{eqnarray}
where, ${\bf n}({\bf r})$ is an O(3) order parameter for the spin,
$\rho_s$ is the spin stiffness, $q$ the deviation in number density of
electrons from uniform filling, $\bar \rho$ is the average density of
electrons, $\nu$ the landau level filling fraction, $\mu$ is a
chemical potential for Skyrmions, and $V({\bf r})$ is the Coulomb
interaction potential. The filling fraction and average electron
number density are related by, $\bar \rho=\nu/2\pi l^2$, where $l$ is
the magnetic length.
Aside from the Coulomb interaction this is 
the usual continuous approximation for the energy of a
ferromagnet. The Coulomb term arises because, in the 
quantum Hall effect, the electric charge density is proportional to
the topological charge density \cite{sondhi}:
\begin{equation}
q = - \frac{\nu}{8\pi}\epsilon_{\mu\nu}\left({\bf
n}.[\partial_{\mu}{\bf n}\times \partial_{\nu}{\bf n}]\right).
\end{equation}

\subsection{Skyrmion Solution.}

It is useful to parametrise the $O(3)$ field, ${\bf n}({\bf r})$, as follows:
\begin{eqnarray}
w = \frac{n_x + \mbox{i}n_y}{1 - n^z}, \: n_x + \mbox{i}n_y =
\frac{2w}{1 + |w|^2}, \: n^z = \frac{|w|^2 - 1}{1 + |w|^2}.\label{linearmap}
\end{eqnarray}
Using this parameterisation of ${\bf n}$, the non-linear sigma
part of the energy functional is,
\begin{eqnarray}
E_{nl\sigma} =  \int \frac{4d^2x}{(1 +
|w|^2)^2}\left[
           \rho_s (|\partial w|^2 +|\bar\partial w|^2)  
          -\frac{\mu}{4\pi}  (|\partial w|^2 - |\bar\partial w|^2)
         \right],
\end{eqnarray}
where $z = x +iy$, $\partial=\partial/\partial z$ and
$\bar\partial=\partial/\partial{\bar z}$. This is minimised on
configurations called instantons or Skyrmions \cite{belavin},
which have the form,
\begin{equation}
w(z) = h\prod_{i=1}^{N}\left(\frac{z - a_i}{z - b_i}\right), \label{w}
\end{equation}
where ${a_i,b_i,h}$ are parameters. $a_i$ and $ b_i$ are called the
coordinates of the instanton. According to ref.[\onlinecite{belavin}], the
energy of such a configuration is proportional to its topological
charge, and does not depend on $\{h, a_i,b_i\}$: $E = (4\pi\rho_s - \mu)N$.

\section{The Effect of Coulomb and Zeeman interactions.}

\subsection{Distortion of Skyrmions.}
For a particular configuration of Skyrmions, the weak Zeeman and Coulomb
interactions cause
the spin distribution to distort slightly in order to minimise the
energy. The energy of these distorted Skyrmions is not degenerate in
their positions. The original distribution, $w=w_0$, becomes
$w=w_0+f$; $w_0(z)$ is analytic and $f(z,\bar z)$ is some non-analytic
function which we assume to be small compared with $w_0$. 

Expanding the variation of the non-linear sigma model part of
the energy to first order in $f$, and using $Q_0=\ln(1+|w_0|^2)$ for
ease of notation, we obtain,
\begin{equation}
\frac{\delta E_{nl\sigma}}{\delta w} =
8\rho_s[2w\partial \bar{w} \bar{\partial} \bar{w}-(1 +
|w|^2)\partial \bar{\partial} \bar{w}]/(1 + |w|^2)^3
=-8\rho_s \bar{\partial}(e^{-2Q_0} \partial \bar{f}).
\end{equation}
The total energy is minimised, that is $\frac{\delta
E_{nl\sigma}}{\delta w}+\frac{\delta (E_{Coulomb}+E_{Zeeman})}{\delta
w}=0$, when,
\begin{eqnarray}
e^{-2Q_0}\partial \bar{f}&=&\int^{\bar z}
d 
\bar{\epsilon}\bar{w_0}(\bar{\epsilon})G(w_0(z),\bar{w_0}(\bar{\epsilon})), \nonumber\\
e^{-2Q_0} \bar{\partial}f &=&\int^{z}
d\epsilon w_0(\epsilon)G(w_0(\epsilon),\bar{w_0}(\bar{z})),  \nonumber\\
G(w(z),\bar{w}(\bar{z}))&=&\frac{1}{8\rho_s}e^{-Q}\frac{\delta
(E_{Coulomb}+E_{Zeeman})}{\delta Q}.\label{f}
\end{eqnarray}
where $Q({\bf x})=\ln [1+|w|^2({\bf x})]$. 
The total energy is then,
\begin{equation}
E[w]=E_{nl\sigma}[w_0]+E_{z+c}[w_0]+E^{ex}_{nl\sigma}[w_0]+E^{ex}_{z+c}[w_0],\label{totenergy}
\end{equation}
where,
\begin{eqnarray}
E^{ex}_{nl\sigma}&=&8\rho_s \int d^2x e^{-2Q}|\bar{\partial}f|^2,\nonumber\\
E^{ex}_{z+c}&=&\int d^2x e^{-Q}(w\bar f+\bar wf)\frac{\delta
E_{z+c}}{\delta Q}.\label{extra1}
\end{eqnarray}
By integrating by parts one may show that
$E^{ex}_{z+c}=-2E^{ex}_{nl\sigma}$. From Eq.(\ref{f}) we see that $f$
is of order $g_L B/\rho_s$ and that the contributions to the total
energy in Eq.(\ref{totenergy}) are of order $\rho_s$, $g_L B$ and
$(g_L B)^2/\rho_s$ respectively. 

Therefore, for small Zeeman
coupling, $\rho_s \gg g_L B$, we may simply ignore these distortions 
(therefore from now on we shall drop the subscript 0), and minimise on the
space of Skyrmion solutions given by Eq.(\ref{w}). 
The energy functional with
these approximations is,
\begin{eqnarray}
E &=& (4\pi\rho_s - \mu)N+E_{Zeeman}+E_{Coulomb},\nonumber\\
E_{Zeeman} &=& -g_L \bar \rho B\int d^2x (1-2e^{-Q}),\nonumber\\
E_{Coulomb} &=& \frac{1}{2}
\left(\frac{e\nu}{4\pi}\right)^2 \int d^2x
\int d^2x' 4 \partial \bar \partial Q(1) V({\bf x}_1-{\bf x}_2) 4
\partial \bar \partial Q(2) \nonumber\\
 &=& 8\left(\frac{e\nu}{4\pi}\right)^2 \int d^2x_1
\int d^2x_2  \nabla^ 2_{1}V({\bf x}_1-{\bf x}_2) Q(1)
\partial \bar \partial Q(2) .\label{energyfunctional}
\end{eqnarray}
In these expressions, $Q(1,2)=Q({\bf x}_{1,2})$ and the notation has been 
relaxed so that $w$ now represents the undistorted Skyrmion spin 
distribution.
We will use this convention from now on.

Taking the functional derivative of Eq.(\ref{energyfunctional}) with 
respect to $w$, we obtain the extremal condition,
\begin{equation}
g_L\bar \rho B e^{-Q}=\left(\frac{e\nu}{\pi}\right)^2 
\int d^2x_2  \nabla^ 2_{1}V({\bf x}_1-{\bf x}_2)
\partial \bar \partial Q(2) .
\end{equation}
We have not been able to solve this equation analytically. It turns out, 
however, that for the experimentally relevant 
range of parameters we can come up with reasonable approximations. 

\section{Approximating the Energy Functional and Its Derivatives.}

\subsection{The Energy Functional.}

In order to make a useful approximation to the energy functional, we 
assume that the spin distribution is very sharply peaked around the 
zeros, $a_i$.
This implies
that the topological charge density at the zeros is much greater than
the average charge density, which is in turn much greater than the
charge density at the poles. That is, $q_{a_i}>>\bar q>>q_{b_i}$.  
Later, we will demonstrate the self-consistency of this approximation for 
the experimental range of parameters. 

For spin distributions of this form, we may approximate the Coulomb
and Zeeman energies by expanding $w$ around its poles
and zeros and integrating the corresponding approximate energy
densities over patches around these points.
The leading order behaviour of $w$ near a pole or zero is,
\begin{eqnarray}
w(z\simeq a_i)=\sqrt{\pi q_{a_i}}(z-a_i),\nonumber\\
1/w(z\simeq b_i)=\sqrt{\pi q_{b_i}}(z-b_i).
\end{eqnarray}
$q_{a_i}$ and $q_{b_i}$ are the topological charge densities at $a_i$
and $b_i$, respectively:
\begin{eqnarray}
\pi q_{a_i}=\frac{|\partial w|^2}{(1+|w|^2)^2}=|\partial w(a_i)|^2
=|h|^2\prod_{j=1,j\ne 
i}^{N}|a_i-a_j|^2/\prod_{j=1}^{N}|a_i-b_j|^2,\nonumber\\
\pi q_{b_i}=\frac{|\partial (1/w)|^2}{(1+1/|w|^2)^2}=|\partial 
[w(a_i)]^{-1}|^2
=\frac{1}{|h|^2}\prod_{j=1,j\ne
i}^{N}|b_i-b_j|^2/\prod_{j=1}^{N}|b_i-a_j|^2.
\end{eqnarray}
The variations of the topological charge density near $a_i$ and $b_i$ are,
\begin{eqnarray}
\pi q(z\simeq a_i)&=&\frac{\pi q_{a_i}}{(1+\pi 
q_{a_i}|z-a_i|^2)^2},\nonumber\\
\pi q(z\simeq b_i)&=&\frac{\pi q_{b_i}}{(1+\pi q_{b_i}|z-b_i|^2)^2}.
\end{eqnarray}
We are now in a position to approximate the Coulomb and Zeeman
contributions to the energy functional. Firstly, the Zeeman energy is
evaluated using the above approximations and integrating over circles
of radius $d$ centred on each $a_i$ and $b_i$. $\pi d^2$ is the area
per half Skyrmion and so $2\pi d^2$ is the area per Skyrmion. The
radius $d$ is related to the average Skyrmion density by, 
\begin{equation}
\bar q =1/2\pi d^2, \: \bar q = |\nu/\nu_0 - 1|\bar\rho.
\end{equation}
>From Eq.(\ref{energyfunctional}), we find,
\begin{eqnarray}
E_{Zeeman}=
\frac{g\bar \rho B}{\bar q}\left( \sum_{a_i} 2\bar
q/q_{a_i}\ln(1+ q_{a_i}/2\bar q)-\sum_{b_i} 2\bar 
q/q_{b_i}\ln(1+q_{b_i}/2\bar
q)\right).\label{Zeeman}
\end{eqnarray}
Secondly, we evaluate the Coulomb energy. From Eq.(\ref{energyfunctional})
we find,
\begin{equation}
E_{Coulomb}=\frac{1}{2}\sum_{i}\left( \frac{e\nu}{\pi} \right)^2
\frac{\sqrt{\pi q_{a_i}}}{\epsilon} 
\underbrace{ \int d^2x_1
d^2x_2\frac{1}{(1+|x_1|^2)^2}\frac{1}{(1+|x_2|^2)^2} }_{\mbox{dimensionless
constant} \sim \mbox{O(1)}.}
+\frac{1}{2}\sum_{i,j}
\frac{(e\nu)^2}{\epsilon |a_i-a_j|}.\label{Coulomb}
\end{equation}
The first term is due to the local charge density at $a_i$ and the
second is an interaction between the charge densities near $a_i$ and
$a_j$. The latter has the form of a point particle interaction
potential, since all the charge is concentrated in small regions of
radius $1/\sqrt{\pi q_{a_i}}$ about $a_i$. The first term is dominant
because, $q_{a_i}>>\bar q \simeq 1/|a_i-a_j|^2$. Note that although
the dominant contribution depends only upon the local density of
charge, this local density depends on the Skyrmion coordinates in a
non-local way, through $q_{a_i}$.

The total energy with these approximations is,
\begin{equation}
E=\frac{g\bar \rho B}{\bar q}\sum_{i}\left( 
2\bar q/q_{a_i}\ln(1+q_{a_i}/2\bar q)- 
2\bar q/q_{b_i}\ln(1+q_{b_i}/2\bar q)
+2\alpha\sqrt{q_{a_i}/2\bar q}\right)
+\frac{1}{2}\sum_{i,j}\frac{(e\nu)^2}{\epsilon |a_i-a_j|},
\end{equation}
where,
\begin{equation}
\alpha = \frac{\bar q}{g_L \bar \rho B} 
\left(\frac{e\nu}{2\pi}\right)^2
\frac{\sqrt{2\pi\bar q}}{\epsilon}
=\frac{\nu_0 |\nu/\nu_0-1|^{3/2}}{(2\pi)^2} \frac{e^2}{\epsilon l g_l B} 
\approx (0.4 \rightarrow 1.6)\nu_0|\nu/\nu_0-1|^{3/2}\label{alpha}.
\end{equation}
In final expression, we have substituted the typical experimental values 
of the ratio $g_L B \epsilon l/e^2 \approx 0.01 \rightarrow 0.04$ 
\cite{barrett,aifer,schmeller}.

\subsection{Derivatives of the Energy Functional.}
In order to determine the minimal Skyrmion configuration and its
stability, we require the first and second derivatives of the energy
functional with respect to the Skyrmion positions. These may be
approximated by evaluating the derivative of the total energy,
Eq.(\ref{energyfunctional}), and expanding the resulting integrals as
in the preceeding Section. 
The derivative of the energy with respect to the
position of a pole or zero, $r_i$, is
\begin{eqnarray}
\frac{\partial E}{\partial r_i}=
&\pm& \frac{g\bar \rho B}{\bar q} \sum_k \left(
 2\bar q/q_{a_k}\ln(1+q_{a_k}/2\bar q)-(1+q_{a_k}/2\bar q)^{-1}
-\alpha\sqrt{q_{a_k}/2\bar q} \right)/(a_k-r_j)\nonumber\\
&\pm& \frac{g\bar \rho B}{\bar q}\sum_i\left(
2\bar q/q_{b_k}\ln(1+q_{b_k}/2\bar q)-(1+q_{b_k}/2\bar q)^{-1}
\right)/(b_k-r_j).\label{firstderiv}
\end{eqnarray}
The plus/minus sign is taken when $r_i$ corresponds to a
zero/pole. The second derivatives are given by,
\begin{eqnarray}
\frac{\partial^2 E}{\partial r_i \partial s_j}=
&\pm& \frac{g\bar \rho B}{\bar q} 
\sum_k \left(
2\bar q/q_{a_k}\ln(1+q_{a_k}/2\bar q)
+(1+q_{a_k}/\bar q)(1+q_{a_k}/2\bar q)^{-1}
+1/2\alpha\sqrt{q_{a_k}/2\bar q}
\right)/(a_k-r_i)(a_k-s_j) \nonumber\\
&\pm& \frac{g\bar \rho B}{\bar q} 
\sum_k -\left(
2\bar q/q_{b_k}\ln(1+q_{b_k}/2\bar q)
+(1+q_{b_k}/\bar q)(1+q_{b_k}/2\bar q)^{-1}
\right)/(b_k-r_i)(b_k-s_j), \label{func}
\end{eqnarray}
where the plus sign is taken when both $r_i$ and $s_j$ correspond to 
a pole or zero and the minus sign otherwise.
For the present we have ignored the contribution from the residual
Coulomb interaction. This will be reintroduced later.

\section{The classical minimum.}
The classical minimum is attained when the derivative of the energy is
zero with respect to all of the Skyrmion coordinates. From
Eq.(\ref{firstderiv}), this occurs when, to logarithmic accuracy,
\begin{eqnarray}
q_{a_i}&=&q_{0}=2\bar q \left( \frac{-2\ln{\alpha}}{3\alpha}
\right)^{2/3},\: 
q_{b_i}\simeq 0,\:
\forall_{i}.\label{constraint}
\end{eqnarray}
As we have already mentioned, our approximation requires $q_0 >> \bar q$ 
which  according to Eq.(\ref{alpha}) is achieved for  a deviation of 
$\nu$ 
from $\nu_0$ of less than $0.2$ ($q_0/\bar q \approx 5\rightarrow 20$ for 
experimental values of $g_L B \epsilon l/e^2 \approx 0.01 \rightarrow 
0.04$ 
and $|\nu/\nu_0-1|=0.2$). 

Eq.(\ref{constraint}) gives $2N$ constraints on the values of 
$q_{a_i}$ and $q_{b_i}$,
and so constrains the positions, ${a_i,b_i}$. For any particular
arrangement of ${a_i}$,we may satisfy the above constraint by fixing
${b_i}$. The number of degrees of freedom is reduced by a factor of
two. At the classical minimum the intuitive picture holds. Equal sized 
Skyrmions with $w=z/r_0$, matched to the ferromagnetic background,
move independently except for the residual Coulomb interaction given
by the second term in Eq.(\ref{Coulomb}). In the single Skyrmion
sector, this is the result of Sondhi {\it et al}. \cite{sondhi}.

 At low enough temperatures Skyrmions interacting in this way are 
expected to form a classical Wigner crystal. The calculation of 
Bonsall {\it et al}. \cite{bonsall}, performed for an electronic 
Wigner crystal, suggests a hexagonal lattice as the minimal solution. 
We argue that the same calculation is valid for the crystal of Skyrmions. 
This brings us into  disagreement with the results of 
Ref.[\onlinecite{brey}].
However, this is ?simply? because these authors did not consider a 
hexagonal 
lattice of the proper type. The simplest periodic arrangement of 
Skyrmions 
is described by a bi-periodic analytic
(elliptic) function, $w$, with two poles and two zeros per unit cell;
$b$, $b'$ and $a$, $a'$ respectively. A general property of such an
elliptic function is that  $|w(z-a)|^2=|w(z-a')|^2$ and
$|w(z-b)|^2=|w(z-b')|^2$. The dependence of the energy functional on
$|w|^2$ and so $n^z$ alone, allows us to treat $a$ and $a'$, (and
also $b$ and $b'$), on an equal footing. The Wigner lattices have
$a$'s and $a'$'s on alternate lattice sites. The hexagonal lattice
predicted here, therefore, is not the higher energy hexagonal lattice
investigated by L. Brey {\it et al}., but rather a distortion of their face
centred square lattice to a face centred rectangular lattice with
sides in the ratio $1:\sqrt{3}$. Figures 1 and 2 show the azimuthal
components of the spin distributions corresponding to the face centred
square and face centred rectangular lattice. Analytic expressions for
these are given by Jacobi elliptic functions; $w=h \;
\hbox{cn}(z,1/\sqrt{2})$ for the square lattice and $w=h \;
\hbox{cn}(z,e^{i2\pi/3})$ for the hexagonal/rectangular lattice.
Since our analysis has neglected dipole and higher moments of the charge 
distribution, the real structure  is likely to be  a rectangular 
lattice with the ratio of sides close to $1:\sqrt{3}$.

To be sure of this result, we must investigate its stability by
evaluating the phonon spectrum. In addition to the usual phonon modes
of the Wigner crystal there will be modes corresponding to
fluctuations in the size of the Skyrmions and to spinwaves. 

\section{Phonon Spectrum of the Skyrme Crystal.}

\subsection{The Wigner crystal.}
Here we present in condensed form the results of Bonsall {\it et al}. for
the normal modes of a classical Wigner crystal. The residual coulomb
interaction is,
\begin{equation}
E=\frac{1}{2}\sum_{i,j}\frac{(e\nu)^2}{\epsilon |a_i-a_j|}
,\label{residcoulomb}
\end{equation}
and its derivatives with respect to the positions of the Skyrmions
are,
\begin{eqnarray}
\frac{\partial^2 E}{\partial a_i \partial a_j} &=& 
-\frac{3(e\nu)^2}{4\epsilon }
\frac{\bar a_{ij}^2}{|a_{ij}|^5} , \nonumber\:
\frac{\partial^2 E}{\partial a_i \partial a_i} =
+\frac{3(e\nu)^2}{4\epsilon } \sum_{j\ne i}
\frac{\bar a_{ij}^2}{|a_{ij}|^5} , \nonumber\\
\frac{\partial^2 E}{\partial a_i \partial \bar a_j} &=& 
-\frac{(e\nu)^2}{4\epsilon } 
\frac{1}{|a_{ij}|^3} , \nonumber\:
\frac{\partial^2 E}{\partial a_i \partial \bar a_i} =
+\frac{(e\nu)^2}{4\epsilon } \sum_{j\ne i}
\frac{1}{|a_{ij}|^3} .
\end{eqnarray}
These are Fourier transformed to give the dynamical matrix,
\begin{eqnarray}
\tilde D_{aa}(k)&=&
\frac{\pi (e\nu)^2 }{2\epsilon} \frac{\bar k^2}{|k|},
\nonumber\\
\tilde D_{a\bar a}(k)&=&
\frac{\pi(e\nu)^2 }{2\epsilon} |k|,
\nonumber\\
D_{rs}(k)&=&\frac{1}{V_c}\sum_{\bf{G}}\tilde D_{rs}(k),
\end{eqnarray}
where $\bf G$ is a reciprocal lattice vector.
To evaluate the normal modes one must first recast these summations into
a more rapidly convergent form, Ref.[\onlinecite{bonsall}]. The result
for the hexagonal crystal, written in terms of the longitudinal and
transverse components, $r,s \in \{a_L,a_T,\}$, is,
\begin{equation}
D^{Wigner}_{rs}=
\left( \begin{array}{cc}
{\cal L}(|{\bf k}|-0.181a_0|{\bf k}|^2)&0\\
0&{\cal T}|{\bf k}|^2
\end{array} \right),
{\cal L}=\frac{2 \pi (e\nu)^2}{\epsilon V_c},
{\cal T}=\frac{2 \pi(e\nu)^2}{\epsilon V_c} 
0.036a_0.\label{Wigner}
\end{equation}
$a_0$ is the lattice spacing and $V_c$ is the area of a unit cell.

\subsection{Contribution of Non-Local Terms.}
For a periodic lattice at the classical minimum, the second
derivatives of the nonlocal part of the energy functional (\ref{func}) 
are given by,
\begin{equation}
\frac{\partial^2 E}{\partial r_i \partial \bar s_j} =
\pm \left( \frac{e \nu}{2\pi} \right)^2
\frac{3\sqrt{\pi q_0}}{2\epsilon} 
\sum_{a_k\ne r_i,s_j} \frac{1}{(a_k-r_i)(\bar a_k-\bar s_j)}
\end{equation}
Terms like $\partial^2 E/\partial a_i \partial a_j$ are zero for a
periodic lattice. Taking the Fourier transform of these derivatives,
we obtain the elements of the dynamical matrix,
\begin{eqnarray}
D_{a \bar a}&=&
3(e\nu)^2\frac{\sqrt{\pi q_0}}{2 \epsilon V_c}
\sum_{{\bf GG}'}
\left(
\frac{1}{(k+G)(\bar k +\bar G')}
-\frac{1}{(k+G)\bar G'}
-\frac{1}{G(\bar k +\bar G')}
+\frac{1}{G\bar G'}
\right), \nonumber\\
D_{a\bar b}&=&
-\;\;\;\;\;\;\;''\;\;\;\;\;\;\;\;
\sum_{{\bf G}{\bf G}'}
\left(
\frac{e^{-i{\bf G}.{\bf b}}}{(k+G)(\bar k +\bar G')}
-\frac{e^{-i{\bf G}.{\bf b}}}{(k+G)\bar G'}
\right), \nonumber\\
D_{b\bar a}&=&
-\;\;\;\;\;\;\;''\;\;\;\;\;\;\;\;
\sum_{{\bf G}{\bf G}'}
\left(
\frac{e^{-i{\bf G}.{\bf b}}}{(k+G)(\bar k +\bar G')}
-\frac{e^{-i{\bf G}.{\bf b}}}{(k+G)\bar G'}
\right), \nonumber\\
D_{b\bar b}&=&
\;\;\;\;\;\;\;\;\;''\;\;\;\;\;\;\;\;
\sum_{{\bf G}{\bf G}'}
\left(
\frac{e^{-i{\bf G}.{\bf b}}}{(k+G)}\frac{e^{+i{\bf G}.{\bf b}}}{(\bar k 
+\bar G')}
\right).
\end{eqnarray}
The summations over the reciprocal lattice vectors, $\bf G$ and ${\bf
G}'$, are only very slowly convergent and must be re-cast into a more
rapidily convergent form. In order to do this, we use a modification
of the Ewald separation method, (see refs.[\onlinecite{ewald}] and
[\onlinecite{bonsall}]). This involves splitting a slowly convergent
real lattice sum into summations over the real lattice and the
reciprocal lattice, both of which are rapidly convergent.
Firstly, we write the elements of the
dynamical matrix in the following way:

\begin{eqnarray}
D_{a\bar a}({\bf k}) &=&
-\left( \frac{e \nu}{2\pi} \right)^2
\frac{3\sqrt{\pi q_0}}{2\epsilon}
\left( \sum_{{\bf a}_l\ne 0}
\frac{e^{i{\bf k}.{\bf a}_l}}{a_l} \right)
\left( \sum_{{\bf a}_l\ne 0}
\frac{e^{i{\bf k}.{\bf a}_l}}{\bar a_l} \right), \label{deriv1}\\
D_{a\bar b}({\bf k}) &=&
+\left( \frac{e \nu}{2\pi} \right)^2
\frac{3\sqrt{\pi q_0}}{2\epsilon}
\left( \sum_{{\bf a}_l\ne 0}
\frac{e^{i{\bf k}.{\bf a}_l}}{a_l} \right)
\left(\sum_{{\bf a}_l}
\frac{e^{i{\bf k}.({\bf a}_l-{\bf b})}}{\bar a_l-\bar b}
\right), \label{deriv2}\\
D_{b\bar a}({\bf k}) &=&
+\left( \frac{e \nu}{2\pi} \right)^2
\frac{3\sqrt{\pi q_0}}{2\epsilon}
\left( \sum_{{\bf a}_l}
\frac{e^{i{\bf k}.({\bf a}_l+{\bf b})}}{a_l+b}
\right)
\left( \sum_{{\bf a}_l\ne 0}
\frac{e^{i{\bf k}.{\bf a}_l}}{\bar a_l} \right), \label{deriv3}\\
D_{b\bar b}({\bf k}) &=&
-\left( \frac{e \nu}{2\pi} \right)^2
\frac{3\sqrt{\pi q_0}}{2\epsilon}
\left( \sum_{{\bf a}_l}
\frac{e^{i{\bf k}.({\bf a}_l+{\bf b})}}{a_l+b}
\right)
\left( \sum_{{\bf a}_l}
\frac{e^{i{\bf k}.({\bf a}_l-{\bf b})}}{\bar a_l-\bar b}
\right).\label{deriv4}
\end{eqnarray}
Next, using the integral representation,
\begin{equation}
\frac{1}{|{\bf x}|^2}=
\int_{0}^{\infty} dt e^{-t|{\bf x}|^2},
\end{equation}
and the two dimensional theta-function transformation,
\begin{equation}
\sum_{{\bf a}_l} \exp \left\{
-t|{\bf x}-{\bf a}_l|^2 + i{\bf k}.{\bf a}_l \right\}=
\frac{\pi}{V_ct} \sum_{\bf G} \exp \left\{
-|{\bf G}-{\bf k}|^2/4t - i{\bf G}.{\bf x} \right\},
\end{equation}
we consider the lattice sums,
\begin{eqnarray}
\sum_{{\bf a}_l\ne 0} \frac{e^{i{\bf k}.{\bf a}_l}}{|{\bf a}_l|^2}
&=&
\sum_{{\bf a}_l\ne 0} \int_{\gamma}^{\infty} dt e^{-t|{\bf a}_l|^2 + i{\bf
k}.{\bf a}_l}
+
\sum_{\bf G} \frac{\pi}{V_c} \int_{0}^{\gamma} \frac{dt}{t}
e^{-|{\bf G}-{\bf k}|^2/4t}
-
\lim_{x \to 0}
\int_{0}^{\gamma} dt e^{-t|{\bf x}|^2 + i{\bf
k}.{\bf x}}\\
\sum_{{\bf a}_l} \frac{e^{i{\bf k}.({\bf a}_l\pm{\bf b})}}{|{\bf
a}_l\pm{\bf b}|^2}
&=&
\sum_{{\bf a}_l} \int_{\gamma}^{\infty} dt e^{-t|{\bf a}_l\pm{\bf b}|^2 + 
i{\bf
k}.({\bf a}_l\pm{\bf b})}
+
\sum_{\bf G} \frac{\pi}{V_c} \int_{0}^{\gamma} \frac{dt}{t}
e^{-|{\bf G}-{\bf k}|^2/4t \mp i({\bf G}+{\bf k}).{\bf b}}.
\end{eqnarray}
In these expressions, $V_c = {\bar q}^{-1}$ is the area of a unit cell 
and $\gamma $
is the Ewald separation parameter. The result is independent of the
value of $\gamma$, which is chosen so that both the lattice and
reciprocal lattice summations on the right hand side are rapidly convergent.
For small wave-vector, ${\bf k}$, the ${\bf G}=0$ term dominates;
\begin{eqnarray}
\sum_{{\bf a}_l\ne 0} \frac{e^{i{\bf k}.{\bf a}_l}}{|{\bf a}_l|^2}
&=&
\frac{\pi}{V_c} \left( -\hbox{constant}-\ln(|{\bf k}|^2/4\gamma
\right),\nonumber\\
\sum_{{\bf a}_l} \frac{e^{i{\bf k}.({\bf a}_l\pm{\bf b})}}{|{\bf
a}_l\pm{\bf b}|^2}
&=&
\frac{\pi\cos({\bf k}.{\bf b})}{V_c} \left( -\hbox{constant}-\ln(|{\bf 
k}|^2/4\gamma
\right).
\end{eqnarray}
We have used the fact that since ${\bf b}$ lies at the centre of the
unit cell, the second summation is independent of its sign. Finally, we
differentiate with respect to $k=k_x+ik_y$ to obtain,
\begin{eqnarray}
\sum_{{\bf a}_l\ne 0} \frac{e^{i{\bf k}.{\bf a}_l}}{a_l}&=&
-2i\frac{\partial}{\partial k} \sum_{{\bf a}_l\ne 0}
\frac{e^{i{\bf k}.{\bf a}_l}}{|{\bf a}_l|^2}=
\frac{2\pi i}{V_c}\frac{1}{k}, \nonumber\\
\sum_{{\bf a}_l} \frac{e^{i{\bf k}.{\bf a}_l}}{a_l\pm b}&=&
-e^{\mp i{\bf k}.{\bf b}}2i\frac{\partial}{\partial k}\sum_{{\bf a}_l}
\frac{e^{i{\bf k}.({\bf a}_l\pm{\bf b})}}{|{\bf a}_l\pm{\bf b}|^2}=
\frac{2\pi i}{V_c}\frac{e^{\mp i{\bf k}.{\bf a}}}{k} .
\end{eqnarray}
Substituting for these expressions into Eqs.(\ref{deriv1}) to 
(\ref{deriv4}), the
contribution to the dynamical matrix from the non-local interaction is,
\begin{equation}
D_{rs}^{non-Local}=
\left( \begin{array}{cccc}
0&0&1&-e^{i{\bf k}.{\bf b}}\\0&0&-e^{-i{\bf k}.{\bf b}}&1\\
1&-e^{i{\bf k}.{\bf b}}&0&0\\-e^{-i{\bf k}.{\bf b}}&1&0&0
\end{array} \right)\delta /|{\bf k}|^2,\label{nonLocal}
\end{equation}
where,
\begin{equation}
\delta=
3\bar q^2 (e \nu )^2
\frac{\sqrt{\pi q_0}}{2\epsilon}. \label{delta}
\end{equation}
This does not depend upon the lattice type. The interaction is so long
range that it is insensitive to the small scale distribution of the
Skyrmions. In fact  our final result is just that obtained using a
continuous Fourier transform of the derivatives. 

\subsection{Phonon Spectrum of the Skyrmion Crystal.}
To detemine the spectrum of the Skyrme crystal, we require the
eigen-modes of the total dynamical matrix,
$D^{Wigner}_{rs}+D^{non-Local}_{rs}$. We will concentrate on the
hexagonal lattice, since this is the minimum energy at our level of 
approximation. In terms of the longitudinal and transverse components, 
$r,s\in \{a_L,b_L,a_T,b_T\}$, the dynamical matrix is,
\begin{equation}
D_{rs}=
\left( \begin{array}{cccc}
{\cal L}|{\bf k}|&0&0&0\\0&0&0&0\\0&0&{\cal T}|{\bf k}|^2&0\\0&0&0&0
\end{array} \right)
+
\left( \begin{array}{cccc}
1&-e^{i{\bf k}.{\bf b}}&0&0\\-e^{-i{\bf k}.{\bf b}}&1&0&0\\
0&0&1&-e^{i{\bf k}.{\bf b}}\\0&0&-e^{-i{\bf k}.{\bf b}}&1
\end{array} \right)2\delta /|{\bf k}|^2
\end{equation}
The first matrix is given by Eq.(\ref{Wigner}). The second matrix is 
obtained from Eq.(\ref{nonLocal}) by multiplying on the
left and right by matrices $\Lambda^T$ and $\Lambda$ respectively,
such that,
\begin{equation}
\left( \begin{array}{c} a_L\\b_L\\a_T\\b_T \end{array} \right)=
\underbrace{\frac{1}{2|{\bf k}|}
\left( \begin{array}{cccc}
k_x&0&k_y&0\\0&k_x&0&k_y\\k_y&0&-k_x&0\\0&k_y&0&-k_x
\end{array} \right)
\left( \begin{array}{cccc}
1&0&1&0\\0&1&0&1\\-i&0&i&0\\0&-i&0&i
\end{array} \right)}_{\Lambda^{-1}}
\left( \begin{array}{c} a\\b\\ \bar a\\ \bar b \end{array} \right)
\end{equation}
The problem separates into two independent sectors. In the
longitudinal sector, the eigen values and eigen-vectors for small
wave-vector are,
\begin{equation}
{\cal L}|{\bf k}|/2,
\left(
\begin{array}{c} 1\\e^{-i{\bf k}.{\bf b}}\end{array}
\right)
\; \hbox{and} \;
4\delta/|{\bf k}|^2,
\left(
\begin{array}{c} 1\\-e^{-i{\bf k}.{\bf b}} \end{array}
\right).
\end{equation}
In the transverse sector the eigen-modes are,
\begin{equation}
{\cal T}|{\bf k}|^2/2,
\left(
\begin{array}{c} 1\\ e^{-i{\bf k}.{\bf b}} \end{array}
\right)
\; \hbox{and} \;
4\delta/|{\bf k}|^2,
\left(
\begin{array}{c} 1\\- e^{-i{\bf k}.{\bf b}} \end{array}
\right).
\end{equation}
The $1/|{\bf k}|^2$ modes are extremely stiff, and basically ensure
that at finite Zeeman coupling the Skyrmions have a fixed size. Note,
however, that according to Eq.(\ref{delta}) $\delta \propto (g_L \ln g_L 
)^{1/3}$ and so at zero Zeeman energy we expect substantial variation in 
the size.
The remaining modes are the usual Wigner crystal modes. 

\subsection{Quantum Fluctuations of Phonons.}
The above calculation of the derivatives of the energy functional 
describes the properties of the Skyrmion crystal in regimes where the 
quantum fluctuations may be neglected. We require a full quantum 
treatment 
of the phonon spectrum. The dynamical part of the action for Skyrmions 
in the quantum Hall effect is given by the Wess-Zumino term,
\begin{equation}
S_{WZ}=
\frac{\nu \bar \rho}{2} \int d^2x dt  {\bf A}[{\bf n}({\bf
r})].\frac{\partial {\bf n}}{\partial t}
=
\frac{\nu \bar \rho}{2} \int d^2x dt d\tau 
{\bf n}.(\partial_t {\bf n} \times \partial_{\tau} {\bf n})
=
2\nu \bar \rho \int d^2x dt d\tau 
\frac{|\partial_T w|^2-|\bar \partial_T w|^2}{(1+|w|^2)^2},
\end{equation}
where $T=t+i\tau, \partial_T=\partial/\partial T$ and 
$\bar \partial_T=\partial/\partial \bar T$. We have extended the function 
${\bf n}(t)$ to a function ${\bf n}(t,\tau)$ in the usual way, such that 
$t$ 
and $0 \le \tau \le 1$ parameterise the interior of the region bounded by 
the 
curve ${\bf n}(t)$ with ${\bf n}(t,1)={\bf n}(t)$ \cite{stone}. 
This 
puts the action into a manifestly gauge invarient form. The third 
expression 
is obtained by substitution of the linear map, Eq.(\ref{linearmap}), into 
this 
gauge invarient form.

We may approximate this contribution to the action in the same way as the 
energy functional, by assuming that the spin distribution is sharply 
peaked 
around the zeros, $a_i$. Integrating over patches around these points we 
obtain,
\begin{eqnarray}
S_{WZ} &=& 
2\nu \bar \rho \int d^2x dt d\tau \sum_i 
\left( |\partial_T a_i|^2 - |\bar \partial_T a_i|^2 \right) 
\pi q_{a_i}({\bf x})
\nonumber\\
&=&
2\pi \nu \bar \rho \sum_i \int dt d\tau 
\left( |\partial_T a_i|^2 - |\bar \partial_T a_i|^2 \right)
\nonumber\\
&=& 
- 2\pi \nu \bar \rho \sum_i \int dt d\tau 
\left(\partial_{\tau} a_{i,x} \partial_t a_{i,y} -\partial_{\tau}
a_{i,y} \partial_t a_{i,x} \right) 
\nonumber\\
&=& 
- \pi \nu \bar \rho \sum_i \int dt 
\left( a_{i,x} \partial_t a_{i,y} -
a_{i,y} \partial_t a_{i,x} \right) 
\nonumber\\
&=& 
\pi \nu \bar \rho  \sum_{i} \int dt 
{\bf a}_i \epsilon  \partial_t{\bf a}_i, 
\end{eqnarray}
where $\epsilon$ is the two dimensional anti-symetric tensor,
\begin{equation}
\epsilon=\left( \begin{array}{cc} 0&-1\\ 1&0 \end{array}  \right).
\end{equation}
This is simply the result presented by M. Stone \cite{stone2} for 
the single Skyrmion and is the Lorentz force for a particle of unit 
charge moving in a magnetic field $2\pi\nu\bar\rho=e\nu^2B$. 
Combining this with the approximation to the energy functional made 
above, we obtain an effective action for the phonon modes of the Skyrmion 
crystal as follows:
\begin{equation}
S_{eff}=\frac{1}{2}\sum_{\omega}\int \frac{d^2k}{(2\pi)^2} 
\delta {\bf a}(\omega,-{\bf k}) 
\left(   i\omega e\nu^2 B \epsilon  - D({\bf k}) 
 \right) 
\delta {\bf a}(\omega,{\bf k}).
\end{equation} 
 Since 
we are only interested in phonon modes which do not change the size of 
the 
Skyrmions, we retain only the corresponding elements of $D_{rs}$. 
Therefore, 
$D({\bf k})$ is the two by two matrix $D^{Wigner}_{rs}/2$ given by 
Eq.(\ref{Wigner}).  
The equation of motion for these fluctuations is
\begin{equation}
\left(  
i \omega e\nu^2 B \epsilon  - D({\bf k}) 
 \right) 
\delta {\bf a}(\omega,{\bf k})=0.
\end{equation}
If we follow Refs.[\onlinecite{stone2,dziarmaga}] and introduce a kinetic 
term, $m^* |\partial_t {\bf n}|^2/2$, into the Lagrangian, the equation 
of 
motion becomes,
\begin{equation}
\left( \omega^2 m^* +i\omega e\nu^2 B  \epsilon  -D({\bf k}) \right)
\delta {\bf a}(\omega,{\bf k})=0.
\end{equation}
Substituting explicit expressions for the matrix $D({\bf k})$ from 
Eq.(\ref{Wigner}) we find that the frequencies of the longitudinal and 
transverse modes are,
\begin{eqnarray}
\omega_t &=& \frac{\sqrt{{\cal LT}}}{2 \omega_c m^*} |{\bf k}|^{3/2},
\nonumber\\
\omega_l &=& \omega_c=\frac{eB}{m^* c}.\label{phononmodes}
\end{eqnarray}
This is identical to the result for an electronic Wigner crystal in a 
magnetic 
field given in Ref.[\onlinecite{fukuyama}].

\section{Spin waves in the Skyrmion Crystal.}
The analysis up to now has been concerned with finding the
groundstate spin distribution. We have argued that this is a slightly
distorted version of the Belavin-Polyakov Skyrmion solution and have
minimised the energy functional on the space of these solutions. When
considering the dynamical properties of the system one must include
the possibility of spin waves. These have been excluded up to now for
a fundamental reason; one cannot write a spin wave in terms of
distortions of the parameters $\{ a_i,b_i\}$. The spinwaves are
non-analytic distortions of the BP Skyrmions. For
example, in the zero charge sector the BP solution is simply the
ferromagnetic ground state and does not allow for spin waves.

Including dynamical terms, the action for Skyrmions in the quantum
Hall effect is \cite{sondhi}
\begin{equation}
S=\int d^2x dt \frac{\nu \bar \rho}{2} {\bf A}[{\bf n}({\bf
r})].\frac{\partial {\bf n}}{\partial t}
-\int dt E,
\end{equation}
where $E$ is given by Eq.(\ref{energy}) and ${\bf A}[{\bf n}({\bf r})]$ is
the vector potential of a unit monopole in spin space. Ignoring the
Coulomb interaction, this gives the following equation of motion for
spin waves \cite{lifshitz,stone}:
\begin{equation}
\frac{\nu \bar \rho}{2} \frac{\partial {\bf n}}{\partial t}
=
\rho_s {\bf n} \times \nabla^2 {\bf n}
+\bar \rho g_L {\bf n}\times {\bf B}.
\end{equation}
These are known as the Lifshitz equations. For small Skyrmions, we expect 
essentially the same spin wave
dispersion as for the ferromagnetic groundstate. In fact, on large length 
scales, the spin distribution looks like a ferromagnet, aside from at
isolated points where the Skyrmions lie. The spectrum is,
\begin{equation}
\omega ({\bf k})=\frac{2}{\nu \bar \rho} 
\left( \rho_s |{\bf k}|^2 +\bar \rho g_L B \right).
\end{equation}
In addition to the gapped ferromagnetic spin wave, there is an
anti-ferromagnetic mode with a linear spectrum, asociated with the
anti-ferromagnetic order of the in plane components of the spin
distribution \cite{sachdev}. In principle one may calculate the spectrum of this mode
by coarse graining the action for small fluctuations about the
Skyrmion crystal on a lattice with half the period of the Skyrmion
crystal. Staggering the effective action on this lattice, taking the
continuum limit and integrating out the fast modes will give an
effective action for the anti-ferromagnetic fluctuations. The solution
of this problem is extremely complicated and depends on all the
parameters of the system in a complicated way through the ground state
configuration. In any case, our primary goal is to discus the melting
of the Skyrmion lattice, a proccess which is unaffected by the
presence of spin waves.

Including the Coulomb interaction will cause a mixing between these
spin waves and the Wigner crystal modes determined above. It is
difficult to estimate this mixing and we will assume it to be negligable.

\section{Melting of the Skyrme Crystal.}
There are three steps in our analysis of the melting of the Skyrmion 
crystal: 
Firstly, we discuss the regime in which quantum fluctuations may be 
neglected.
Next, we describe the various possibilities for defect mediated
melting of the crystal and compare the predictions with
experiment. Finally, we estimate at what temperature
significant fluctuations in the size of the Skyrmions are to be expected.

\subsection{Quantum Fluctuations.}
In the usual way, quantum fluctuations of a particular mode may be 
neglected 
if the temperature is sufficiently high that the first Matsubara 
frequency, 
$2\pi T$, is larger than the energy of the mode at the momentum cut-off. 
With a high momentum cut-off at the first Brillouin zone, 
$|{\bf k}_{max}|^2=\pi^2/a_0^2=\sqrt{3} \pi^2\bar q/2$, using the phonon 
dispertion from 
Eq.(\ref{phononmodes}) and substituting the expressions for ${\cal L}$ 
and ${\cal T}$ from Eq.(\ref{Wigner}), we obtain 
the condition,
\begin{equation}
T \agt  0.03\frac{(e\nu_0)^2}{\epsilon l}
 |\nu/\nu_0-1|^{3/2}
\sim 5 |\nu/\nu_0-1|^{3/2} K, \label{fluct}
\end{equation}
for the temperature above which quantum fluctuations of the phonon mode 
may 
be neglected. As indicated, for experimental values of the paramenters 
this 
temperature is only a few tenths of a Kelvin. 
A similar analysis for  
spinwaves performed in Ref.[\onlinecite{green}] gives the following 
estimate 
for the temperature above which these quantum fluctuations can be neglected:
\begin{equation}
T \agt \frac{2\rho_s}{\nu^2}|\nu/\nu_0-1|, \label{fluct1}
\end{equation}
which for $\nu_0$ = 1 gives the temperature of order of 
$10|\nu/\nu_0-1|K$. 
If conditions (\ref{fluct}, \ref{fluct1}) are satisfied, 
the Skyrmion crystal becomes  an entirely classical object.  
In fact it can be called a crystal only in a loose sense. Since the 
second derivatives of energy  are proportional to  $|{\bf k}|^2$, the 
fluctuations of displacements diverge logarithmically :
\begin{equation}
<u^2>=\frac{V_c T}{(2\pi)^2} \int \frac{d^2k}{{\cal T}|{\bf k}|^2} 
\sim \frac{V_c T}{{\cal T}} \ln (L \sqrt{\bar q}),
\end{equation}
where L is the system size. In these circumstances, there is no long 
range 
positional order. As usual in two dimensions,
however, one may have a finite stiffness, ${\cal T} \ne 0 $ which is 
associated with a directional long range order. 

\subsection{2-Dimensional Melting.}
The melting transition in two dimensions is driven by the presence of 
defects;  dislocations and disclinations \cite{Kleinert}. First
attempts at explaining 
the melting transition treated these two types of defect as independent.
The resulting system displays two second order Kosterllitz-Thouless type 
transitions, as first the dislocations and then the disclinations unbind 
with increasing temperature. 
At the dislocation unbinding transition we pass from the crystal
phase, with long range directional order and power law decay of
positional order, to a liquid crystal phase which has exponential
decay of potitional order and a power law decay of orientational
order, (for an underlying hexagonal symetry, this phase is known as a
hexatic). The second transition is from the hexatic to an isotropic
liquid which has exponential decays in both positional and
orientational order. 
This picture, the Kostelitz-Thouless-Nelson-Halperin-Young (KTNHY)
\cite{nelson} model, has several 
successes. Allowing for renormalisations of the elasticities due to the 
presence of phonons and dislocations, the lower temperature transition 
from the solid to the hexatic phase occurs at the same temperature as seen 
in some experimental systems, (notably the electronic Wigner crystal), and 
numerical simulations. Unfortunately, this is not the full story. Even those 
melting transitions whose temperatures are predicted correctly show some 
characteristics of being first order. Singularities in specific heat are 
always too sharp to be adequately described by a Kosterlitz-Thouless type 
transition.

A mechanism by which defect mediated melting in two dimensions could be 
first order was pointed out by Kleinert \cite{Kleinert1}. He noted that
in fact dislocations  
and disclinations are not independent. A dislocation can be viewed as a 
bound pair of disclinations and a disclination as a string of dislocations. 
Although the formation of an isolated disclination has an enormous cost in 
strain energy, this is screened by the presence of dislocations. In fact 
the dislocation density acts like an effective temperature for the 
disclinations, and virtual fluctuations can lead to a Kosterlitz-Thouless 
transition at which disclinations proliferate. The presence of disclinations 
also modifies the effective potential of the dislocation density, which 
develops a second minimum at some non-zero density of dislocations. For a 
certain temperature this minimum is at zero energy and the feedback between 
the dislocations and disclinations drives a first order transition directly 
to the liquid. This occurs at a temperature below the Kostelitz-Thouless 
transition temperature. This picture is still only a caracature of
the actual melting process. In fact, dislocations need not pile up on
neighbouring sites to form a disclinations. If dislocations form a
string along next nearest neighbour sites, the resulting configuration
is identical to a grain boundary and the transition may be driven
first order by a proliferation of these boundaries. It is sufficient
that one allows for the possibility of dislocations forming strings by
not choosing a prohibitive form of the core energy \cite{Kleinert1}.

The question remains which of these models is applicable to a
particular system. A lattice model of defects 
demonstrating all of these features has been presented by Kleinnert
\cite{Kleinert2}.
Whether the melting proceeds via two second order transitions or a
single first order transition depends upon the length scale of local
rotational stiffness. The following partition function describes a
lattice with dislocation and disclination defects:
\begin{eqnarray}
Z &=& \prod_x \left( \int^{\pi}_{-\pi} du_i(x) \right)
\sum_{ \{n_{ij}(x),m_i(x) \}} 
\exp (-\beta E), \nonumber\\
E &=& \frac{1}{4} \sum_{x,ij} 
\left( \nabla_i u_i +\nabla_j u_i - 2\pi n_{ij} \right) ^2
+\frac{2\lambda}{\mu} \sum_{x,i} 
\left( \nabla_i u_i - \pi n_{ii} \right)^2
+\frac{2l^2}{a^2} \sum_{x,i} \left(\nabla_i w -2\pi m_i \right)^2
\label{strainenergy}
\end{eqnarray}
The particles are placed at the sites of a simple square lattice with
sites labeled, $x$. Derivatives are all lattice derivatives; $\nabla
f(x)=f(x+i)-f(x),\bar \nabla_i f(x)=f(x)-f(x-i)$. $u_i(x)$ is the
displacement in the $i$ direction of the particle at site $x$, 
normalised so that when $u_i=2\pi$, the
atom has been displaced by one lattice period, $a$. The inverse
temperature is, $\beta=a^2\mu/(2\pi)^2T$. With this normalisation,
$\lambda$ and $\mu={\cal T} V_c/2$ are the usual Lam\'{e} elastic
constants. Fure future reference, we define the Poisson ratio as,
$\tilde \nu = \lambda/(\lambda+2 \mu)$. Summations over integer $n_{ij}$ and even $n_{ii}$ allow for
periodic jumps of atoms to neighbouring sites. This models the
presence of dislocations. In the absence of dislocations, the first
two terms appearing in the energy functional are simply the energy
given by linear elasticity theory. The presence of disclinations is
acounted for by the third term. Disclinations are very singular
objects which cause large distortions of the lattice. In the vicinity
of the disclination linear elasticity theory breaks down, and one must
include a term which is proportional to the square of the local
rotation field, $w=(\nabla_1 u_2-\nabla_2 u_1)/2$. $l$ characterises
the length scale over which the lattice is stiff to local rotations.

One final point is that the sum over $n_{ij}$ is unconstrained so that
in effect atoms can hop onto neighbouring sites with no energy
cost. For molecular or electronic crystals, one must modify the theory
to allow for the singular repulsion between particles. For the Skyrme
crystal, this is not necessary since two Skyrmions may easily sit on
top of one another to form a doubly charged Skyrmion.

A dual representation of this model in terms of stress gauge fields,
may be obtained, by introducing the stress tensor, $\sigma_{ij}$, and
the rotational stress tensor, $\tau_i$, via a
Gaussian transformation and then integrating out the strain fields,
$u_i(x)$ and $w(x)$. The stress gauge fields are defined in terms of
the stress tensors as, 
$\sigma_{ij} = \epsilon_{ij} \bar \nabla_k A_k,
\tau_i = \epsilon_{ij} \bar \nabla_j h - A_i$.
In terms of these, the energy functional takes the form,
\begin{eqnarray}
\beta E &=& \sum_{x} \left\{
\frac{1}{4\beta} \left[ 
\frac{1}{1+\tilde \nu} (\nabla_i A_j)^2
-\frac{1}{2} \frac{1-\tilde \nu}{1+\tilde \nu} (\bar \nabla_i A_i)^2 \right]
+\frac{a^2}{8\beta l^2} (\bar \nabla_k h-\epsilon_{kl}A_l)^2
-2\pi i (A_i \bar b_i + h \bar \theta)
\right\},\label{stressgaugefields}
\end{eqnarray}
where $\bar b_i= \epsilon_{jk} \nabla_j n_{ki} - m_i$ and $\bar
\theta= \epsilon_{ki} \nabla_k m_i$ are the integer valued dislocation
and disclination densities respectively.

This partition function may be written purely in terms of the defect
fields, $\bar b_i$ and $\bar \theta$, by integrating out the stress
gauge fields, $A_i$ and $h$. From Eq.(\ref{stressgaugefields}), the
effective partition function in its pure defect form is,
\begin{eqnarray}
Z=\sum_{ \{ \bar b_i(x)\} }
\exp &-& \left[  4\pi^2 \beta (\tilde \nu+1) 
\sum_x \epsilon_{ij} \nabla_i \bar b_j(x) 
(-\bar {\bf \nabla}.{\bf \nabla} )^{-1} 
\epsilon_{kl} \nabla_l \bar b_k(x) \right. \nonumber\\ 
&+& \left.  8\pi^2 \beta  \sum_x \nabla_i \bar b_i(x) 
\left\{ -\bar {\bf \nabla}.{\bf \nabla}  \left(
\frac{a^2}{l^2}-\bar {\bf \nabla}.{\bf \nabla} \right) \right\}^{-1}
\nabla_j \bar b_j(x)
\right] .\label{Zlargel1}
\end{eqnarray}
The character of the phase transitions in this model depend strongly
upon the length scale of rotational stiffness. For very large $l$ the
KTHNY picture holds and there are two continuous transitions.
For infinite $l$ and $\tilde \nu=1$, the partition function reduces to,
\begin{equation}
Z=\sum_{ \{ \bar b_i(x)\} }
\exp \left[ -  8\pi^2 \beta \sum_x \bar b_i(x) (-\bar {\bf
\nabla}.{\bf \nabla} )^{-1} 
\bar b_i(x) \right],
\end{equation}
which has a Kostelitz-Thouless type phase transition at
$4\beta=2/\pi$. For very large but finite $l$, the longitudinal part of the field $\bar
b$, is massive and does not contribute to the critical behaviour. The
remaining contribution, from the transverse component, is given by the
first term in the exponent of Eq.(\ref{Zlargel1}). This may be written,
\begin{equation}
Z=\sum_{ \{ \bar b_i(x)\} }
\exp \left[ -  4\pi^2 \beta (\tilde \nu+1) 
\sum_x \bar b_i(x) \frac{
\delta_{ij} \bar {\bf \nabla}.{\bf \nabla} 
-\nabla_i \bar \nabla_j }{(\bar {\bf \nabla}.{\bf \nabla})^2}
\bar b_j(x) \right].\label{Zlargel2}
\end{equation}
The long range form of the effective potential between the
dislocations, is given by,
\begin{eqnarray}
\frac{ \delta_{ij} \bar {\bf \nabla}.{\bf \nabla} 
- \nabla_i \bar \nabla_j }{(\bar {\bf \nabla}.{\bf \nabla})^2}
& \rightarrow& \lim_{\delta \rightarrow 0} 
\int \frac{d^2q}{(2\pi)^2}
\frac{e^{i {\bf q}.{\bf x}}}{(q^2+\delta^2)^2}
(\delta_{ij}q^2-q_i q_j) \approx 
\frac{1}{4\pi} \left( \delta_{ij}\ln |x|
-\frac{x_i x_j}{|x|^2} \right).
\end{eqnarray}
We have used differentials of the series expansions of the
zeroth order Bessel function in order to obtain this results. 
The pairwise interaction energy of the dislocations is,
$E_{int}=2\pi\beta (1+\tilde \nu) \ln |x|$,
and there is an unbinding transition at a temperature,
$(1+\tilde \nu)\beta=2/\pi, \hbox{ or, } T=a^2 \mu (1+\tilde\nu)/8\pi$.
This is the universal temperature for the dislocation unbinding
transition, predicted by KTNHY. In Ref.[\onlinecite{morf}], Morf
calculated the renormalisation of the elastic constant, $\mu$, for
the electronic Wigner crystal at
finite temperature, due to the presence of phonons and
dislocations. With a suitable rescaling of the
parameters, this result may be applied directly to the Skyrmion
crystal. The melting temperature, allowing for this renormalisation, is
\begin{equation}
T_m=|\nu/\nu_0-1|^{1/2}\frac{(e\nu_0)^2}{\epsilon l} \frac{1}{2 \times
128.2\sqrt{2}} \simeq 0.44 |\nu/\nu_0-1|^{1/2}.
\end{equation}
For small $\beta \approx a^2/l^2$, the effective partition function
is,
\begin{equation}
Z=\sum_{ \{\bar \theta (x) \} }
\exp \left( 2 \frac{l^2}{a^2} 4\pi^2 \beta \sum_x \bar \theta (x)
(-\bar {\bf \nabla}.{\bf \nabla} )^{-1} \bar \theta (x) \right).
\end{equation}
This model is equivalent to a Coulomb gas and has a KT transition at
$4\beta l^2/a^2 \approx 2/\pi$, where the disclinations unbind.

For very small $l$, there is a single first order phase
transition. The constraint $\bar \nabla_kh=\epsilon_{kl}A_l$ 
is forced upon the stress gauge fields, (see
Eq.(\ref{stressgaugefields})). The energy functional then becomes,
\begin{equation}
\beta E =\sum_x \left[ \frac{1}{4\beta(1+\tilde \nu)} (\nabla \bar \nabla h)^2
+2\pi i (\bar b_i \epsilon_{ij} \bar \nabla_i h + \bar \theta
h)
\right].
\end{equation}
This model shows a single first order phase transition at
$(1+\tilde \nu)\beta=0.941$. If one studies the same transition in the model
in which quadratic terms in the original energy functional,
Eq.(\ref{strainenergy}), are replaced by cosines, the predicted
temperature is, $(1+\tilde \nu)\beta=1.328$.
Not allowing for renormalisation of the elastic constants at finite
temperature, {\it i.e.} setting $\mu$ and $\tilde \nu$ to their zero
temperature values, we obtain the following prediction for the melting
temperature:
\begin{equation}
T_m=\frac{(1+\tilde \nu)}{1.328} \frac{a^2
\mu}{(2\pi)^2}=0.48|\nu/\nu_0-1|^{1/2}.
\end{equation}
Allowing for the renormalisation of $\mu$ by phonons at finite
temperature, ($\tilde \nu$ remains virtually
unchanged), this temperature is reduced to
$T_m=0.38|\nu/\nu_0-1|^{1/2}$. For the cosine model the corresponding
temperature is, $T_m=0.34|\nu/\nu_0-1|^{1/2}$ without renormalisation
and $T_m=0.27|\nu/\nu_0-1|^{1/2}$ with \cite{Kleinert3}. Figure 3 shows
a sketch of the way in which the renormalisation curve for $\mu$
calculated by Morf\cite{morf} is used to obtain these corrections to
the transition temperatures. there will of course be additional
renormalisation of the elastic constant which we have not allowed for
here, due to the presence defects.

At some intermediate value of $l$, there is a change from the first order
transition to two second order transitions. On the basis of numerical
simulations to determine the position of this change over, Kleinert
has predicted \cite{Kleinert4} that the Wigner crystal should undergo
a first order transition. The actual value of $l$ for the Wigner
crystal lies very close to the change over, and so the transition is
nearly continuous. We expect that since the Skyrmion crystal shares
the same phonon spectrum as the Wigner crystal, it too should undergo
a first order melting transition.

\subsection{Comparison with Experiment.}
Recent experimental measurements of the heat capacity of a two
dimensional electron system in the quantum Hall regime , made by Bayot
{\it et al.} \cite{bayot} show a marked peak at a temperature of
around $40 mK$ with a  deviation from total filling $\nu_0=1$ of around
$|\nu/\nu_0-1| \sim 0.2$. These authors argue that this contribution
to the specific heat comes from coupling of the nuclear spins to the
Skyrmion spin distribution and that the sharp peak indicates a melting
of the Skyrmion lattice. The sharpness of the peak suggests a first
order transition, which accords with our prediction based on the
criterion for the local rotational stiffness. The temperature of the
first order transition in our lattice defect model is a factor of
three or so greater than the experimental value at $|\nu/\nu_0-1|\sim
0.2$. There are two points to note in reconciling this difference:
(i.)$|\nu/\nu_0-1|=0.2$ is actually outside the regime where quantum
fluctuations of phonons can be neglected at the melting point. These
fluctuations may lead to a reduction in the melting
temperature. (ii.) Softening of the pairwise interaction between
Skyrmions as they begin to overlap means that in regions of large
distortion such as near a defect, the effective lattice stiffness is
reduced and the melting temperature is lowered accordingly.

This latter point may be better seen by comparing our partition
function with that for an N-body system with a pairwise interaction
potential, $\Phi (x-y)$ \cite{Kleinert5}. The partition function is,
\begin{equation}
Z=\int \frac{dx_1...dx_N}{N!} \exp\left[ -\frac{\beta}{2}\sum_{i\ne j}
\Phi({\bf x}_i-{\bf x}_j) \right].
\end{equation}
We consider a crystalline state at low temperature, where the
particles fluctuate around lattice sites, ${\bf x}$, with small
displacements, ${\bf u}$. the partition function may then be written
as,
\begin{equation}
Z=\frac{1}{N!} \sum_{x,i} \prod_{x,i} \left(
\int_{-\infty}^{\infty}du_i({\bf x}) \right)
\exp\left[ -\frac{\beta}{2}\sum_{x\ne y}
\Phi({\bf x}-{\bf y}+{\bf u}({\bf x})-{\bf u}({\bf y})) \right].
\end{equation}
Taylor expanding the partition function about the equilibrium
configuration,
\begin{equation}
Z=\frac{1}{N!} \sum_{x,i} \prod_{x,i} \left(
\int_{-\infty}^{\infty}du_i({\bf x}) \right)
\exp\left[ -\frac{\beta}{2}\frac{1}{4} \sum_{x\ne y}
\partial_i \partial_j \Phi(x-y) (x-y)_k (x-y)_l 
\nabla_k u_i(x) \nabla_l u_j(y) \right].
\end{equation}
For equilibrium configurations including defects, the lattice gradients $\nabla_i
u_j$ are no longer zero, but equal to the jump numbers,
$n_{ij}$. Therefore, we make a similar Taylor expansion in powers of
$\nabla_i u_j -n_{ij}$. It is obvious from this final form of the
partition function and by comparison with Eq.(\ref{strainenergy}) that
a reduction in the curvature of the pair interaction potential
effectively reduces the elastic constant. In going from the Villain,
(quadratic), approximation to the cosine approximation for the elastic
energy, the non-linearities in the cosine allow for such a reduction
in the elastic constant, hence the lower melting temperature.

We conclude that the peak in heat capacity seen by Bayot {\it et al.}
is consistent with a first order defect mediated melting of the
Skyrmion lattice to a liquid of equal sized Skyrmions. Up to
$|\nu/\nu_0-1|\sim0.05$, the melting temperature is expected to vary
as $|\nu/\nu_0-1|^{1/2}$. Beyond this quantum fluctuations are
important and the melting is no longer classical. We sketch this
behaviour in figure 4.

\subsection{Fluctuations in the Skyrmion size.}
Now we discuss the criteria for fluctuations in the size of the Skyrmions 
to be important. The mean square size of these fluctuations is  
\begin{eqnarray}
<u^2>=\frac{V_c T}{8\delta} \int \frac{|{\bf k}|^2 d^2k}{(2\pi)^2}
\sim \frac{V_c T |{\bf k}_{max}|^4}{64 \pi \delta}.
\end{eqnarray}
To estimate when 
these fluctuations are significant, we use a Lindeman criterion; 
$<u^2>/V_c \sim 1$. This is satisfied for temperatures greater than
\begin{equation}
T \agt \frac{128}{\pi^3} \frac{(e\nu_0)^2}{\epsilon l}  |\nu/\nu_0-1|^{1/2}
\left( \frac{-2 \ln \alpha}{3 \alpha} \right)^{1/3}
\sim 700   |\nu/\nu_0-1|^{1/2}
\left( \frac{-2 \ln \alpha}{3 \alpha} \right)^{1/3} K.
\end{equation}
This temperature is always several thousand times the melting 
temperature and in the experimental range of parameters with 
$|\nu/\nu_0-1|\sim 0.2$ is around $300K$. In fact to acheive a realistic 
temperature for this transition, say of the order of $1K$, would require 
a reduction in the effective Zeeman coupling by a factor of $\sim 100$. 
This takes us outside of the regime where our approximations are
self-consistent and so we must be cautious about such
assertions. However, we may be certain that no such fluctuations in
the size of Skyrmions occur in the experimental range of parameters
and that the transition to the meron liquid described in
Ref.[\onlinecite{green}] does not occur. 
 
At temperatures of around $4\pi \rho_s$ significant thermal excitation of 
Skyrmion anti-Skyrmion pairs will begin to occur, and our restriction to 
distorted BP solutions will be inadequate.

\section{Summary.}

At moderate Zeeman energies, the distortion of Skyrmions in the
fractional quantum Hall effect is small and one may describe spin
textures using the Belavin-Polyakov solutions. 
We show that there is a range of parameters where one may 
self-consistently 
assume that the charge density of the spin distribution occurs at 
isolated 
points. This range is spanned by that realised experimentally. 
The spin distribution of many Skyrmions depends in a very non-local way 
upon the Skyrmion coordinates. However, the interplay of Coulomb and 
Zeeman 
energies is such that near the classical minimum the behaviour accords 
with 
the intuitive picture of equal sized Skyrmions interacting via a residual
point-like Coulomb interaction. At low temperatures these form a Wigner 
crystal with hexagonal symetry. When one takes note of the azimuthal 
components of the spn texture, the full hexagonal symetry is no longer 
apparent; the Skyrmions form a face centred rectangular lattice with 
sides close to  the ratio $1:\sqrt{3}$. The analytic expression for 
this dstribution is given by $w=h c\!n(z,e^{2\pi i/3})$. In fact,
accounting for the higher moments in the Coulomb interaction, the
actual symetry is expected to be rectangular with sides in a ration
somewhere between $1:\sqrt{3}$ and $1:1$.

The phonon spectrum of the Skyrmion crystal has four modes. Two of these 
correspond to the phonon modes of the electronic Wigner crystal and two 
extremely stiff modes correspond to fluctuations in the size of the 
Skyrmions. Quantum fluctuations of these modes may be neglected for 
filling 
fractions close enough to $\nu_0=1$ and for a sufficiently high 
temperature. 
We have investigated the defect mediated melting of the Skyrmion
crystal in this classical regime. A first order transition is
predicted which is consistent with recent experiment.
Fluctuations of the size of Skyrmions and the transition 
to a meron liquid requires a huge decrease in the effective Zeeman
coupling and is not expected in the experimental range of parameters.

\section{Acknowledgements.}
We would like to thank J. Chalker and A. Rutenberg for helpful 
discussions and for their interest in the work.

\begin{figure}
\vskip -1.5in
\centerline{\psfig{figure=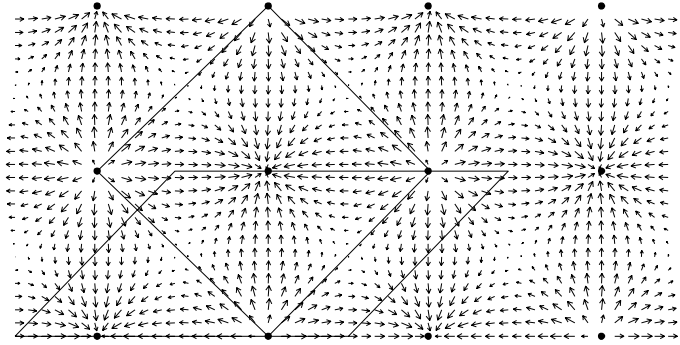,width=5.0in}}
\vskip -0.5in
\caption{{\bf Square} {\bf Skyrmion} {\bf  Lattice:} plot of $(n_x,n_y)$ 
derived from
$w(z)=h \;  c\!n(z,1/ \protect{\sqrt{ 2 })}$. The dots indicate the
zeros of the
function $w(z)$. The square cell is that chosen by Brey et al. and the
parallelepiped is the standard cell.}
\label{squarelatt}
\end{figure}
\begin{figure}
\vskip -1.5in
\centerline{\psfig{figure=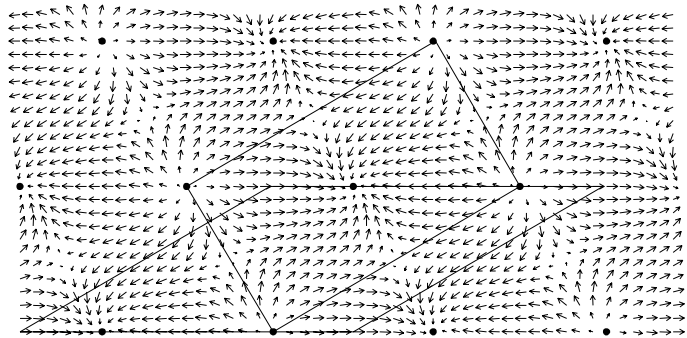,width=5.0in}}
\vskip -0.5in
\caption{{\bf Hexagonal} {\bf  Skyrmion} {\bf Lattice:} plot of $(n_x,n_y)$ 
derived from
$w(z)=h \; c\!n(z,e^{i2\pi/3})$. The dots indicate the zeros of the
function $w(z)$. The solid lines indicate the standard cell and the
face centred rectangular cell.}
\label{trianglatt}
\end{figure}
\begin{figure}
\centerline{\psfig{figure=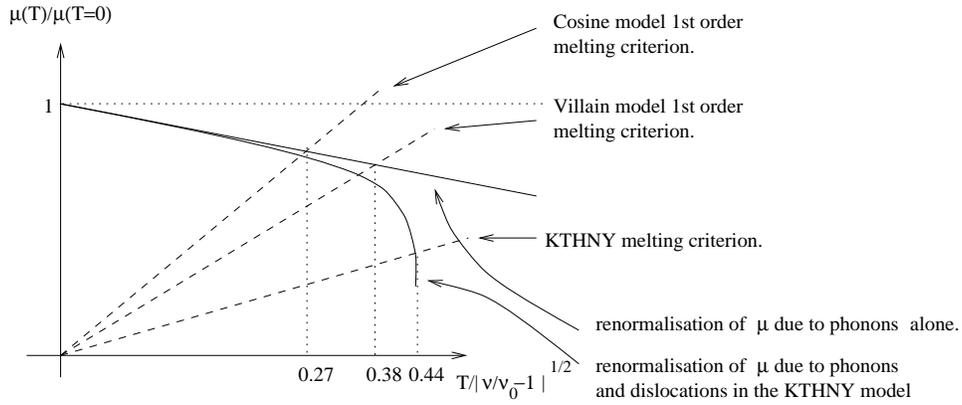,width=5.0in}}
\vskip 0.5cm
\caption{{\bf RG} {\bf flow} {\bf of } {\bf elasticity:} The solid
curve shows the finite temperature renormalisation of the elastic
constant $\mu$ due to the presence of phonons and dislocations. The
straight dashed lines show the various melting criteria. The points of
intersection between the RG curve and the melting criteria give the
melting points. Intersections with the horizontal dotted line at $\mu
(T)/\mu(0)$ give the melting points without renormalisation
corrections.  see Ref.[23]}
\label{RGflow}
\end{figure}
\begin{figure}
\centerline{\psfig{figure=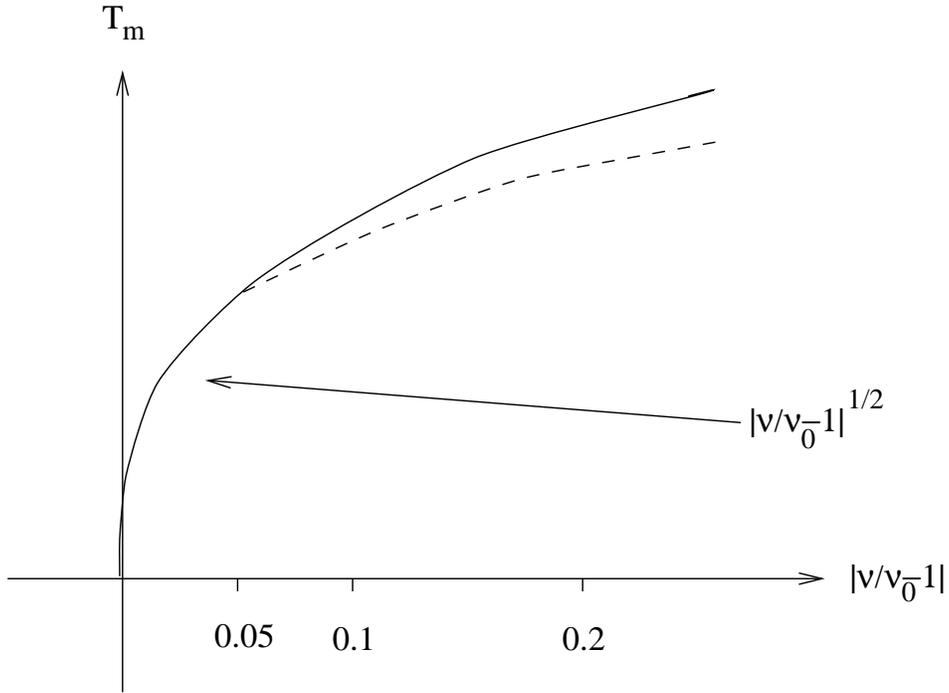,width=5.0in}}
\vskip 0.5cm
\caption{{\bf Predicted } {\bf melting } {\bf temperature } {\bf
versus } {\bf $|\nu/\nu_0-1|$:} Up to $|\nu/\nu_0-1| \sim 0.05$ the system is
classical up to the melting temperature and we expect $T_M\sim
|\nu/\nu_0-1|^{1/2}$. Beyond here quantum fluctuations may lower the
melting temperature.}
\label{Meltingtemp}
\end{figure}


\begin{references}
\bibitem{Kleinert3} In fact the melting temperature predicted for the
Villain and cosine models on the square lattice are
$(1+\nu)\beta_m=0.815$ and $1.15$ respectively. The results for the
triangular lattice are obtained by multiplying by an additional factor
of $2/\sqrt{3}$. see Ch. 14.4 Ref.[\onlinecite{Kleinert}]
\end{references}
\end{document}